\documentclass{aa}
\usepackage{natbib}
\usepackage{psfig}
\usepackage{txfonts}

\newcommand\eg{{\it e.g.} }

\newcommand\etal{et~al.}
\font\aipsfont = cmsy8 scaled\magstep1

\newcommand\aips {{\aipsfont AIPS}}

\begin{document}

\title{A  sensitive  search  for  CO  J=1--0  emission  in  4C\,41.17:
high-excitation molecular gas at $z$=3.8}

\titlerunning{CO J=1-0 in 4C\,41.17}

\author{Padeli P.\ Papadopoulos \inst{1}
\and
Thomas  R.\ Greve \inst{2}
\and 
R.\ J.\ Ivison \inst{3,4}
\and 
Carlos  De Breuck \inst{5}}
\authorrunning{Papadopoulos et al. }
\institute{{Institut f\"ur Astronomie, ETH, 8093 Z\"urich,
 Switzerland} \and 
{California Institute of Technology, Pasadena, CA 91125, USA} \and
 {Astronomy Technology Centre, Royal Observatory, Blackford Hill,
Edinburgh EH9 3HJ, UK} \and
{Institute for Astronomy, University of Edinburgh, Blackford Hill, 
Edinburgh EH9 3HJ, UK} \and
{European Southern Observatory, Karl Schwarzschild Stra\ss e 2,
D-85748 Garching, Germany}}

\offprints{Padeli P. Papadopoulos,
\email{papadop@phys.ethz.ch}}
\date{Received / Accepted}

\abstract{We report  sensitive imaging  observations of the  CO J=1--0
line  emission in  the powerful  high-redshift radio  galaxy 4C\,41.17
($z=\rm 3.8$) with the NRAO Very Large Array (VLA), conducted in order
to detect the large  concomitant H$_2$ gas reservoir recently unveiled
in  this  system  by  \citet{deb05}  via  the  emission  of  the  high
excitation J=4--3  line.  Our observations  fail to detect  the J=1--0
line but yield sensitive  lower limits on the $\rm R_{43}=(4-3)/(1-0)$
brightness temperature ratio of  $\rm R_{43}\sim 0.55-\ga 1.0$ for the
{\it bulk} of the H$_2$ gas mass.  Such high ratios are typical of the
high-excitation molecular gas phase  ``fueling'' the star formation in
local  starbursts,  but  quite  unlike  these  objects,  much  of  the
molecular  gas in  4C\,41.17 seems  to be  in such  a state,  and thus
participating in the observed  starburst episode.  The widely observed
and  unique association  of  highly excited  molecular  gas with  star
forming  sites allows  CO  line emission  with large  (high-J)/(low-J)
intensity ratios  to serve as  an excellent ``marker'' of  the spatial
distribution  of star formation  in distant  dust-obscured starbursts,
unaffected by extinction.

}

\maketitle

\section{Introduction}

The  powerful  radio  galaxy  4C\,41.17  at $\rm  z=\rm  3.8$  with  a
far-infrared  (FIR)  luminosity of  $\rm  L_{\rm FIR}\sim\rm  10^{13}\
L_{\odot}$ is the  site of an enormous starburst  event in the distant
Universe \citep[\eg][]{eal93,gra94}, an interpretation put forth since
its  discovery \citep{cha90}.  This  view received  additional support
from the discovery  of a large dust mass  \citep{chi94,dun94}, and its
concomitant molecular  gas reservoir  with $\rm M(\rm  H_2)\rm \sim\rm
5\times 10^{10}\ M_{\odot}$  \citep{deb05}, which presumably fuels the
star formation.  This follows the  path of similar discoveries made in
the last  decade, uncovering vast  molecular gas masses residing  in a
variety    of    high-redshift    objects.    These    include    QSOs
\citep{bar94,omo96,ohta96},           submm-selected          galaxies
\citep{fra98,neri03},  Ly-break  galaxies  \citep{bak04} and  powerful
radio  galaxies \cite{kla05}.  The  highest redshift  at which  CO has
been detected stands  now at $z=\rm 6.42$, when  the Universe was less
than a Gyr old \citep{wal03,ber03}.

Aside from  the intrinsic  interest of studying  the molecular  gas in
distant  powerful  radio  galaxies,  the  likely  formation  sites  of
present-day   giant  ellipticals   and   their  surrounding   clusters
\citep[\eg][]{pen97,arc01}, the case of  4C\,41.17 is special since it
was  the  first  {\it   non-lensed}  high-redshift  object  for  which
dedicated,  sensitive  observations   were  conducted  to  detect  its
molecular gas mass \citep{ivi96,bar96}.  Observations of CO J=1--0 are
of    particular   importance   since    they   provide    the   least
excitation-biased  method  of   tracing  metal-rich  H$_2$  gas  ($\rm
E_1/k_B\sim 5.5\ K$, $\rm n_{cr}\sim 400\ cm^{-3}$), while most of the
reported searches/detections of molecular gas in high-redshift objects
use CO J+1$\rightarrow $J, J+1$\geq $3  and usually J = 4--3 or higher
transitions   \citep[\eg][]{eva96,ojik97,dow99}.   These  transitions,
with their much higher critical  densities and energy levels (e.g. for
CO 4--3:  $\rm n_{\rm cr}\sim\rm  2\times 10^4\ cm^{-3}$,  $\rm E_{\rm
4}/k_B\sim\rm 55\,K$) trace  much warmer and denser gas,  and thus may
not  reveal the distribution  of the  bulk of the molecular~gas.

Evidence for  molecular gas, sub-thermally excited  over large scales,
has    emerged    for    a    few   objects    at    high    redshifts
\citep[\eg][]{pap02,gre03}  and for  this  reason we  followed up  the
recent  detection  of  CO  J   =  4--3  in  4C\,41.17  with  sensitive
observations of the J = 1--0  transition. At $z=\rm 3.8$ the latter is
redshifted to  24\,GHz and becomes accessible to  the K-band receivers
of  the VLA.\footnote{The  National Radio  Astronomy Observatory  is a
facility of the National Science Foundation operated under cooperative
agreement   by   Associated   Universities,   Inc.}    Our   sensitive
observations, which  unlike past efforts had the  advantage of knowing
the line center from the detection of CO J = 4--3, do not detect any J
=  1--0  emission   but  the  lower  limits  they   set  on  the  $\rm
R_{43}=(4-3)/(1-0)$ line ratio imply a high excitation for the bulk of
the  molecular gas.   This suggests  that, unlike  local counterparts,
most  of   the  molecular  gas   reservoir  in  this   distant  galaxy
participates in the observed starburst event.  Throughout this work we
assume   $\rm   H_{\circ}=75\   km\   s^{-1}\   Mpc^{-1}$   and   $\rm
q_{\circ}=1/2$.

\section{Observations}

The observations  of 4C\,41.17  were conducted with  the VLA in  its D
configuration with four $\sim $8-hr tracks obtained during 2004 August
12--22 UT  and a phase-tracking center  at: RA (J2000)  = 06$^{\rm h}$
50$^{\rm m}$ 52$^{\rm s}$.150 and  Dec (J2000) = 41$^{\circ} $ 30$^{'}
$  30$  ^{''}$.801.   The CO  J  =  1--0  line ($\nu  _{\rm  rest}=\rm
115.2712$\,GHz)  at the redshift  $z=\rm 3.7978$  of the  radio galaxy
\citep[centered on its CO 4--3 emission;][]{deb05} shifts to $\nu_{\rm
obs}=\rm  24.0258$\,GHz,  accessible to  the  VLA's K-band  receivers.
Aiming solely at the detection  of line emission rather than obtaining
information about  the velocity distribution of the  molecular gas, we
used the  continuum correlator mode with  one IF pair  (left and right
circular polarization)  tuned to observe the  line+continuum while the
other one simultaneously observed  the adjacent continuum.  At 24\,GHz
the effective IF bandwidth, $\rm \Delta \nu _{IF}=45$ MHz, corresponds
to $\rm \Delta  V_{IF}\sim 560\ km\ s^{-1}$.  The  wide velocity range
($\sim\rm   1000$\,km\,s$^{-1}$)  revealed   by  the   CO  J   =  4--3
observations ensures  no sensitivity loss due to  velocity dilution of
the J  = 1--0  line within  the IF bands.   Two separate  tunings were
necessary  (and were  equally distributed  per track)  to  observe the
entire  expected  line  width.   These [on-line,  continuum]  tunings,
optimized to observe  the 'blue' and the 'red'  velocity components of
the CO  J = 4--3 emission, were:[24.0649,  24.1649]\,GHz and [24.0351,
23.9351]\,GHz,  corresponding to  [$-490$,  $-1735$]\,km\,s$^{-1}$ and
[$-115$, $+1130$]\,km\,s$^{-1}$ velocity offsets relative to $\nu_{\rm
obs}(z)$.

During  our observing  sessions the  system temperatures  were $T_{\rm
sys}\sim\rm  (70-90)$\,K,  and  a  fast-switching technique  was  used
\citep{car99},  recording  data  every  5\,sec (integration  time  per
visibility) with  20\,sec spent on the  compact calibrator 06465+44513
and 150\,sec  on source; the  typical slewing time to  the calibrator,
3.5$ ^{\circ}$  away, was 20\,sec.   This ensured very  good amplitude
and  phase calibration  with a  resulting  phase $|  \phi _{\rm  rms}|
\la\rm 25^{\circ}$  for most  of the time.   The latter  was estimated
from  the  residual phases  of  our  amplitude/phase calibrator  after
applying antenna  gain solutions smoothed to $\rm  \sim 3\times \Delta
t_{cal}$   ($\rm  \Delta   t_{cal}\sim  3.2\,min$:   interval  between
successive observations  of the  calibrator).  The flux  density scale
was  set  using  3C\,286  (2.59\,Jy)  and  0713+438  (0.55\,Jy).   The
resulting  bootstrapped  flux  of  06465+44513,  used  for  the  final
amplitude/phase  calibration  of the  data,  was  $S(\rm  24\, GHz)  =
3.15\,Jy$ (averaged over all four days), and carries an uncertainty of
$\sim\rm 15$\%.

\subsection{Data reduction: the search for CO J = 1--0 line emission}

\begin{figure}[t]
\psfig{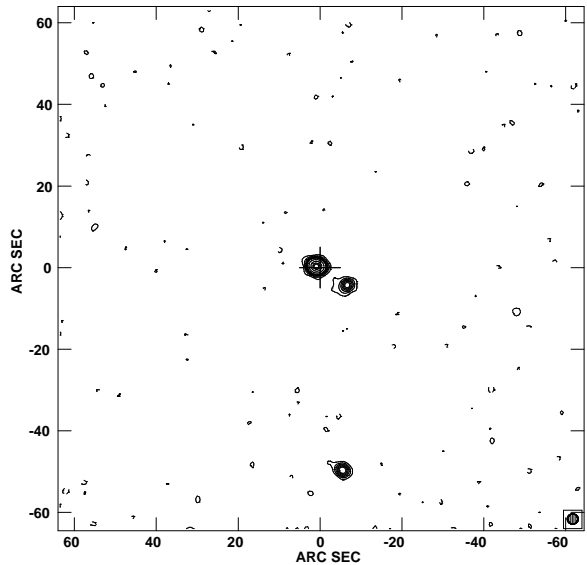} 
\vspace{-0.5cm}
\caption{Radio continuum  map of  4C\,41.17 at 24\,GHz  resulting from
imaging all the $uv$ datasets (see text). Robust weighting was applied
(with {\sc robust} = 0) and  {\sc clean} was employed, converging to a
total flux $S_{\rm CL}=\rm 4.88$\,mJy).  The map was then convolved to
circular  synthesized  beam of  FWHM=$2.5^{''}$  shown  at the  bottom
right. The contours are at $\rm (-3,  3, 6, 9, 12, 15, 18, 30, 50, 70,
80)\times \sigma  $, with  $\rm \sigma =  28\ \mu Jy\  beam^{-1}$. The
cross marks the  position of the radio core, RA  (J2000): 06 50 52.15,
Dec  (J2000): +41  30 30.80  \citep{car94},  which is  also the  phase
tracking center of the observations presented.}
\label{fullfov}
\end{figure}

The measured  visibilities were carefully edited  and calibrated using
standard  techniques described in  the NRAO  \aips\ Cookbook  and then
Fourier transformed  (using the task  {\tt imagr}) to  produce images.
CLEANing was applied to produce  the final (deconvolved) images of the
emission of 4C\,41.17 at  24\,GHz.  In all our naturally-weighted maps
the measured r.m.s.\ noise was close to the theoretical value expected
from the  number of the  visibilities imaged, the  system temperatures
during our observations, two IFs  per frequency and an assumed antenna
aperture  efficiency  of $\eta_a  =\rm  0.40$ at  24\,GHz\footnote{VLA
Observational Status Summary 2004}. A  map of the entire field of view
of the  primary beam ($\sim  2^{'}$), using all  the data is  shown in
Fig.~\ref{fullfov}. The  flux densities measured for the  core and its
SW feature  are: $S_{\rm core}  =\rm (2.97\pm 0.45)\,mJy$  and $S_{\rm
SW}=\rm (0.73\pm  0.12)\,mJy$, in  good agreement with  those reported
for these components by  \citet{ivi96}.  The SW feature corresponds to
feature 'A' in low-frequency maps \citep{cha90,car94} and is dominated
by optically  thin synchrotron  emission from the  radio lobe,  with a
steep   spectral   index   $\alpha   =\rm  -1.7$   ($S_{\nu   }\propto
\nu^{\alpha}$) \citep{ivi96}.  For the emission feature at $\rm \Delta
\theta \sim -50^{''}$  to the South (also seen in  the maps of Carilli
et al.\ 1994)  we find a probability of it due  to chance of $P(\Delta
\theta    )\sim\rm    0.25$,    thus    the   association    is    not
significant.\footnote{We  estimate  $P(\Delta  \theta)$ following  the
method  outlined  by  \citet{dow86}  and  the  1.4-GHz  number  counts
reported  by \citet{hop03}  after  we extrapolate  a  flux of  $S_{\rm
1.4GHz}=\rm  5$\,mJy for  our  source (assuming  a  spectral index  of
$\alpha =\rm -0.7$ between 1.4 and 24\,GHz).}

\begin{figure}[t]
\psfig{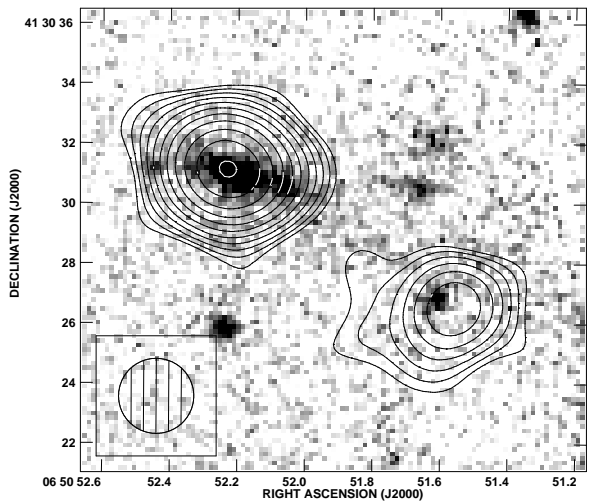} 
\caption{Radio  continuum  map  of  4C\,41.17  at  24\,GHz  (contours)
overlayed on the  2.15-$\mu $m image obtained at  the Keck Observatory
(Graham et  al.\ 1994). The contour  levels are (3, 4.25,  6, 8.5, 12,
17,  24, 34, 48,  68, 96)  $\times \sigma$,  with $\sigma  $=26\, $\mu
$Jy\, beam$^{-1}$.}
\label{KK}
\end{figure}

An overlay of our new 24-GHz map with those obtained at 1.4 and 5\,GHz
\citep{car94}  reveals   excellent  correspondence  of   the  emission
features across all  three wavelengths, a testimony to  the good phase
calibration  at 24\,GHz  resulting from  our fast-switching  scheme. A
deep  $K_S$-band  (2.15\,$\mu$m)  image  shows a  faint  feature  just
upstream from the SW radio  lobe A (see Fig.~\ref{KK}).  At a redshift
of  $z=\rm  3.8$,  2.15\,$\mu$m  corresponds to  rest-frame  $B$-band,
expected  to be  dominated by  young stars.   This is  especially true
since the  $K_S$-band is free of emission-line  contamination, and the
scattered AGN light  is expected to be negligible  $\sim $35\,kpc from
the radio core position.  If  this component is indeed at the redshift
of 4C\.41.17,  it would  indicate a second  locus of  jet-induced star
formation in this distant radio galaxy \citep{dey97,bic00}.

\begin{figure*}[t]
\begin{center}
\begin{tabular}{cc}
\psfig{file=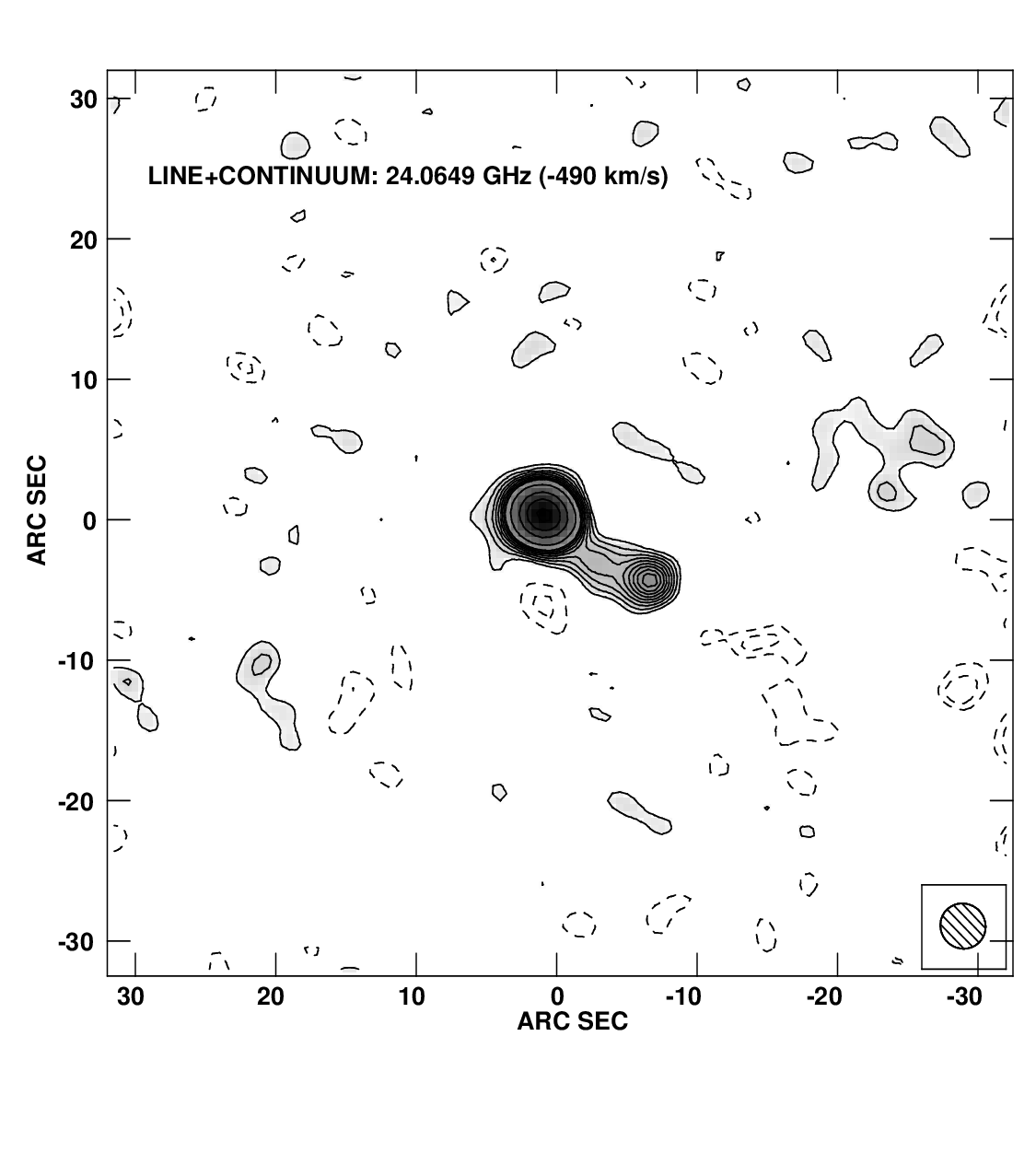,width=8cm}&
\psfig{file=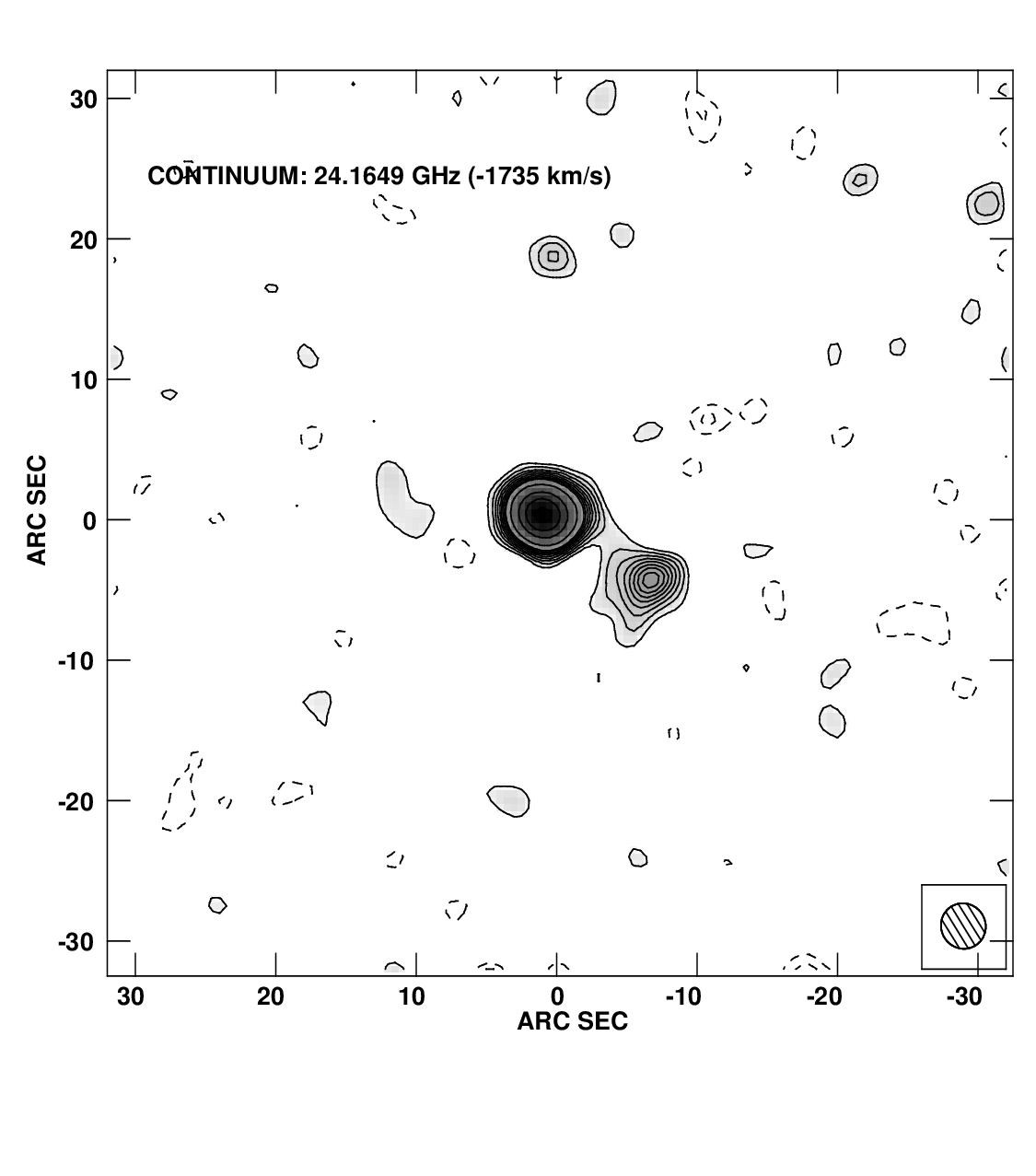,width=8cm}\\
\vspace*{-2.0cm}
\end{tabular}
\psfig{file=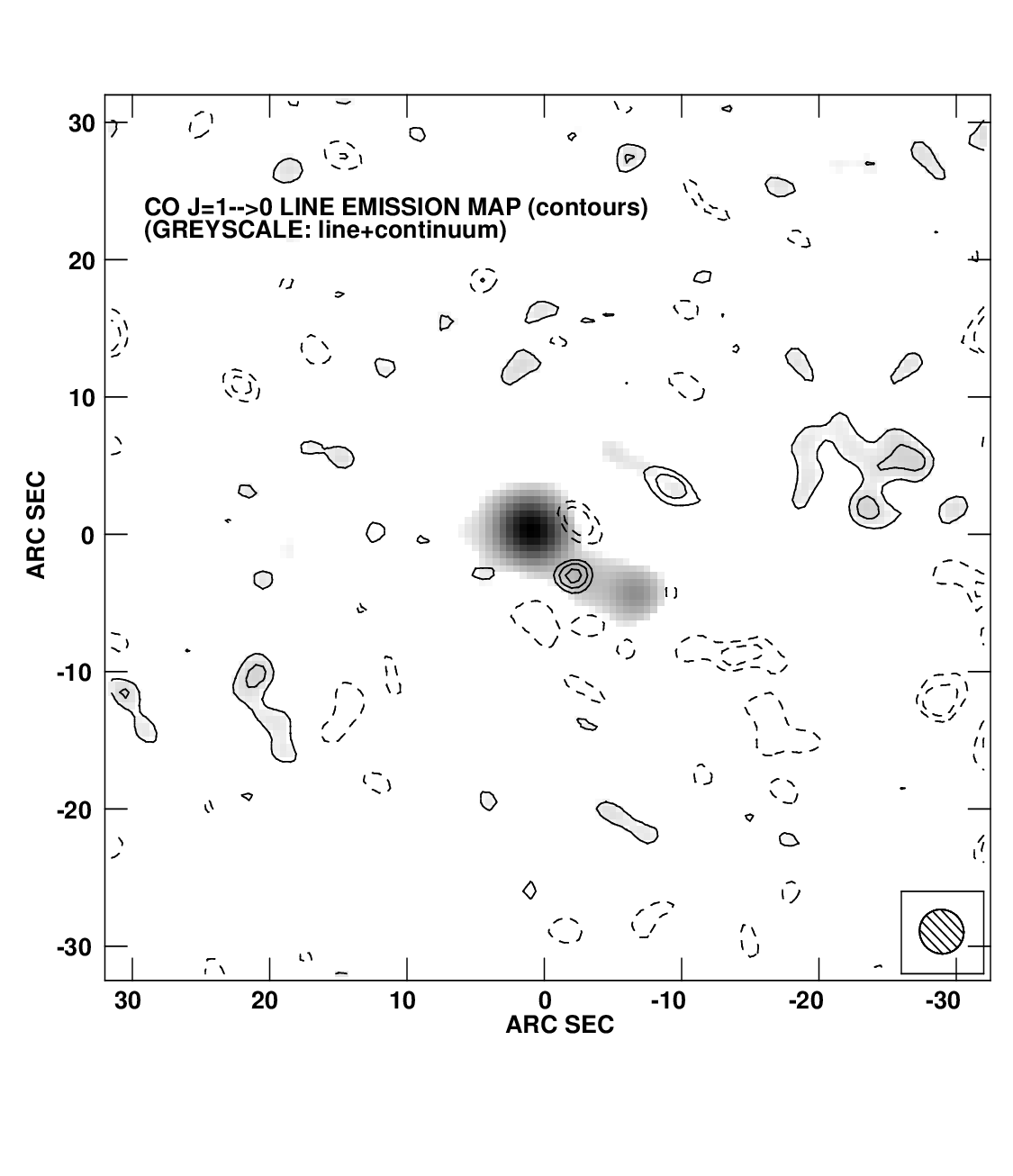,width=15cm} 
\end{center}
\vspace*{-2.5cm}
\caption{Naturally-weighted  maps  of  (``blue'' line)+continuum  (top
left),  continuum (top right),  and line  only (bottom)  emission (see
text). Contours: $\rm (-3, -2, 2, 3, 4, 5, 6, 7, 8, 9, 10, 15, 25, 35,
45,  55,  65,  75, 85)  \times  \sigma$  ($\rm  \sigma  = 55  \mu  Jy\
beam^{-1}$). The restoring beam is $\rm 3.27^{''}\times 3.16^{''}$ and
is shown  at the bottom right  corner ({\sc clean} was  not applied on
the continuum-free map).}
\label{COblue}
\end{figure*}

\begin{figure*}[t]
\begin{center}
\begin{tabular}{cc}
\psfig{file=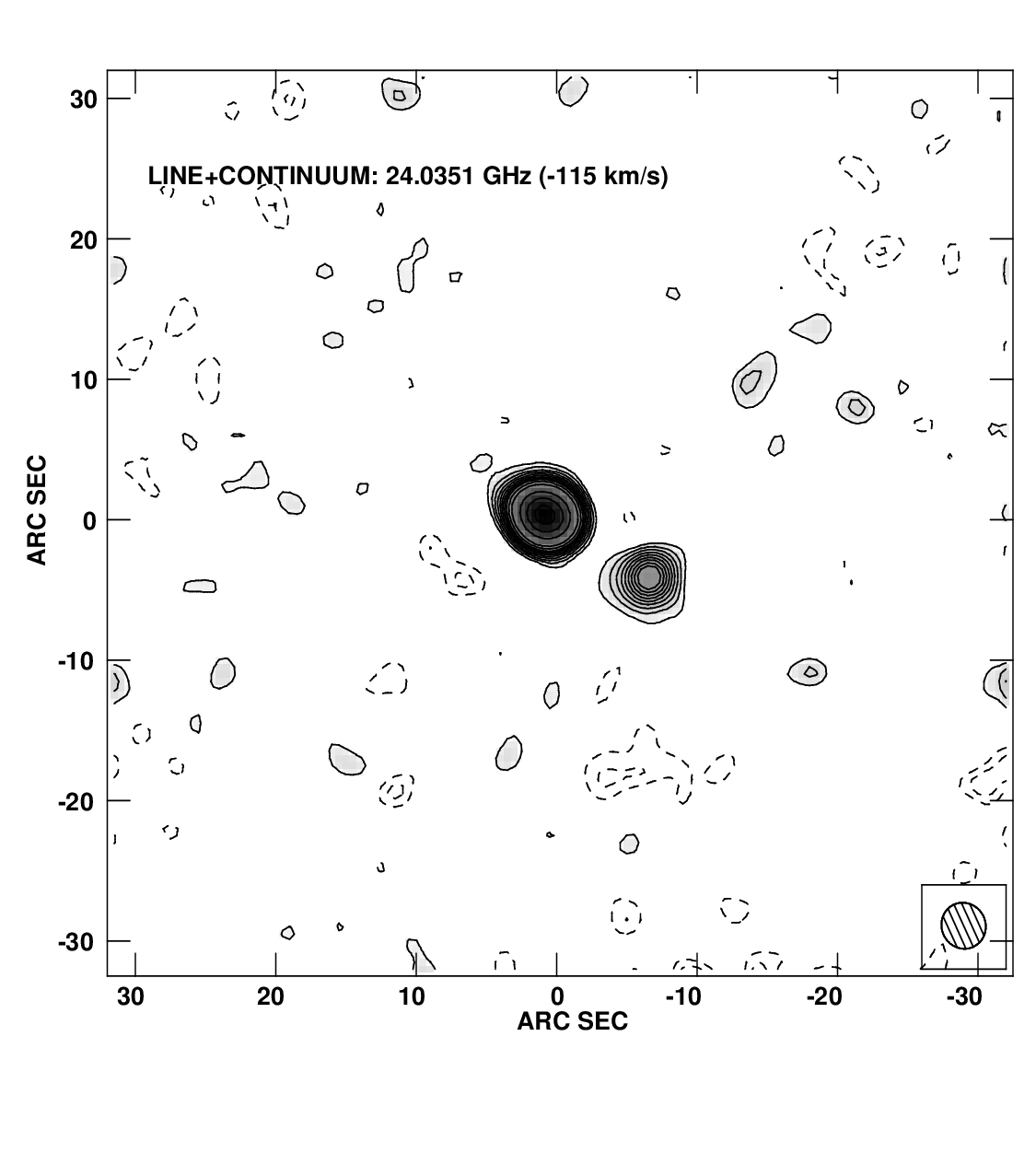,width=8cm}&
\psfig{file=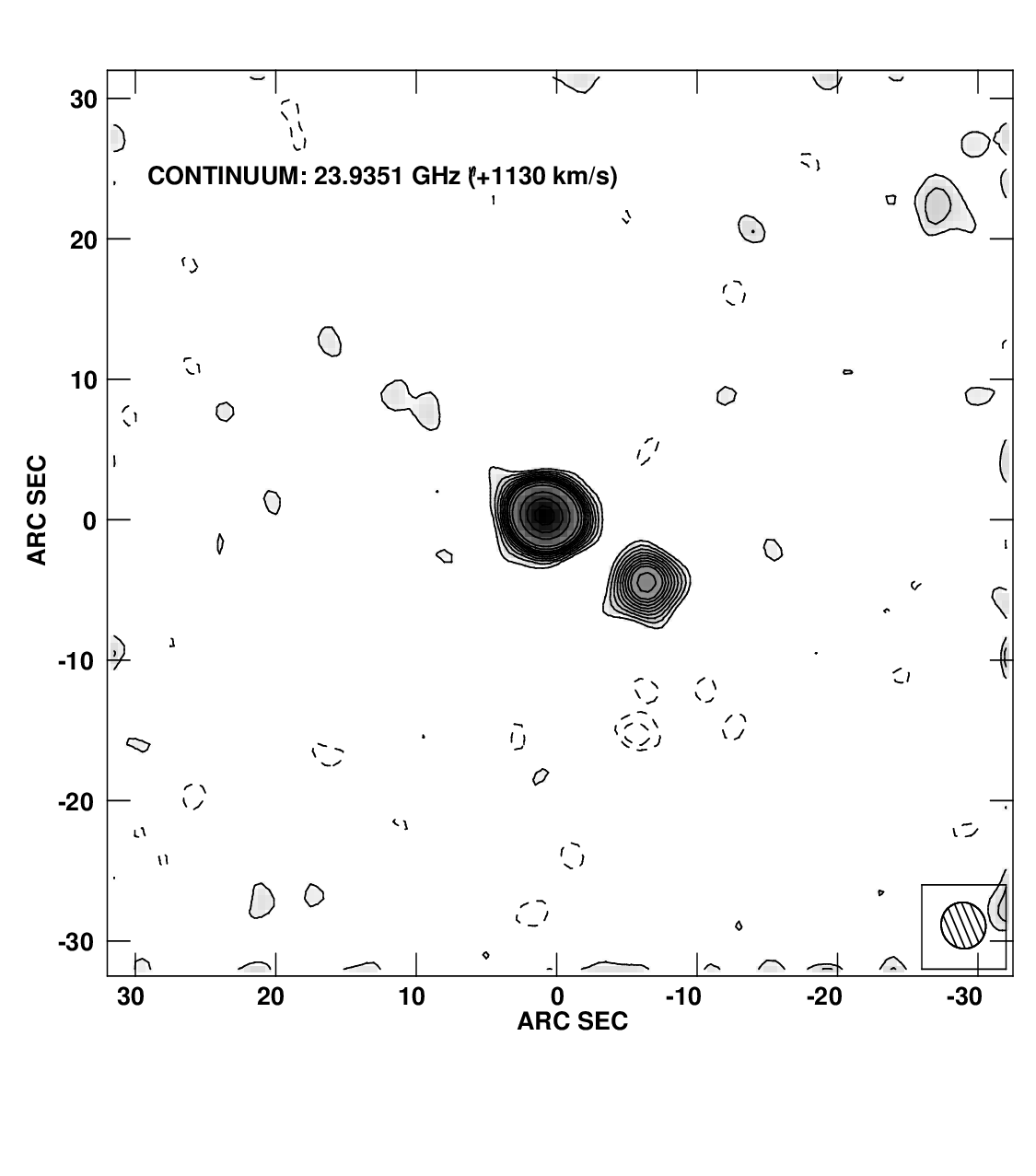,width=8cm}\\
\vspace*{-2.0cm}
\end{tabular}
\psfig{file=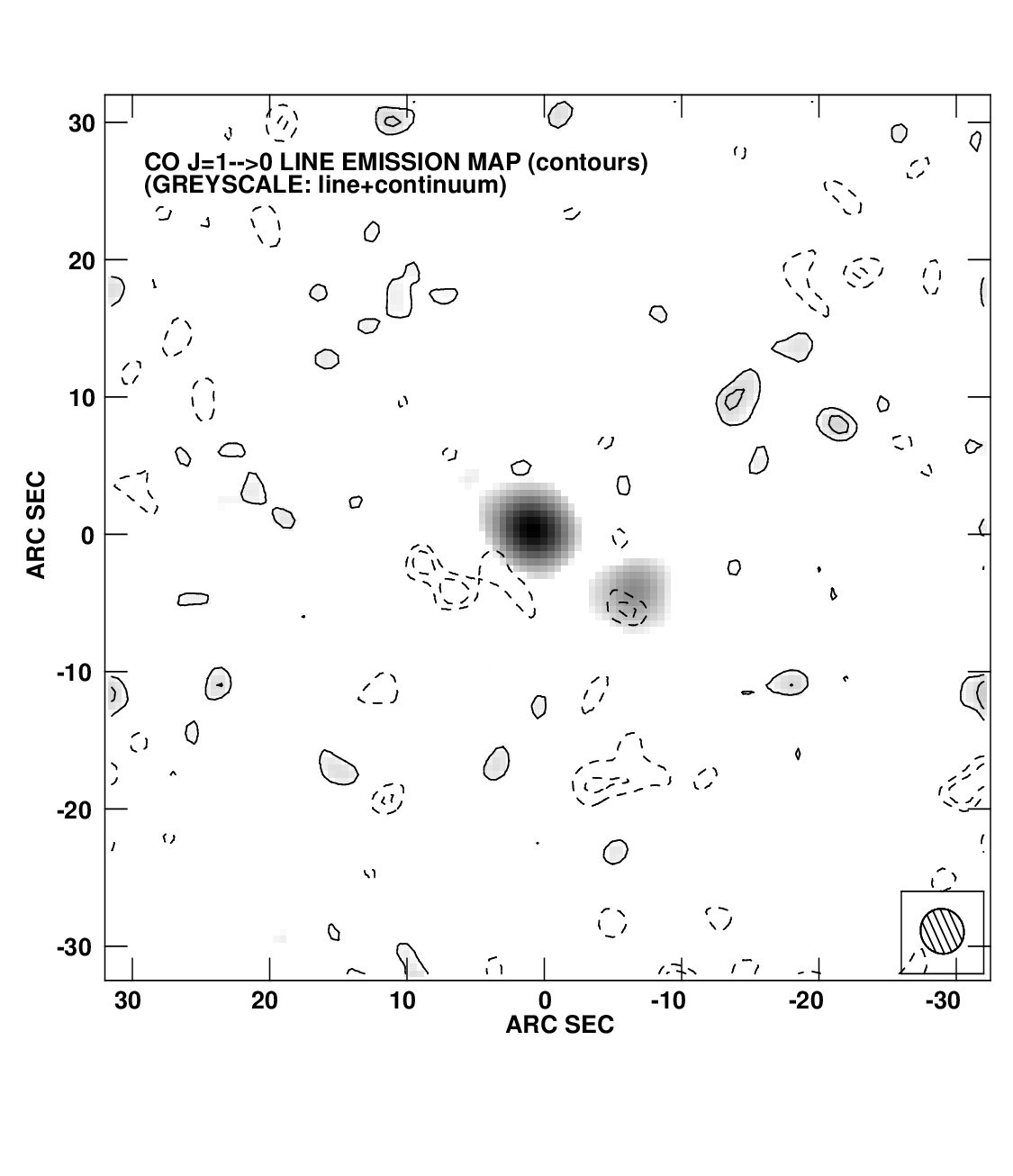,width=15cm} 
\end{center}
\vspace*{-2.5cm}
\caption{Naturally-weighted  maps  of  (``red'' line)+continuum  (top
left),  continuum (top right),  and line  only (bottom)  emission (see
text). Contours: $\rm (-3, -2, 2, 3, 4, 5, 6, 7, 8, 9, 10, 12, 15, 25,
35,  45, 55, 65,  75, 85)  \times \sigma$  ($\rm \sigma  = 55  \mu Jy\
beam^{-1}$).   The restoring beam  is $\rm  3.31^{''}\times 3.13^{''}$
and is shown  at the bottom right corner ({\sc  clean} was not applied
on the continuum-free map).}
\label{COred}
\end{figure*}

We searched  for CO J =  1--0 emission by comparing  the images ``on''
and ``off'' the expected line for the two different frequency settings
covering its velocity range (as revealed by the CO J = 4--3 emission).
We did not find any significant excess flux density in the maps ``on''
with  respect to  those  ``off''  the line,  within  the noise  levels
achieved, while the total CLEANed  flux density of the radio galaxy is
similar for these  maps and for both frequency  settings.  We produced
line-only  maps using  the \aips\  task {\tt  uvsub}: the  {\tt clean}
components of the continuum maps  were Fourier transformed back to the
visibility plane, re-sampled by  the $uv$ sampling function (identical
for the  simultaneously obtained on-line and  off-line datasets), then
subtracted from  the (line  + continuum) visibilities.   The resulting
continuum-free  visibilities were  then  Fourier transformed  (without
CLEANing)  to  produce naturally  weighted  (for maximum  sensitivity)
images  (Figs~\ref{COblue}  and  \ref{COred}).  Significantly  tapered
images  from these  visibilities  were also  made  to investigate  the
presence  of  any  extended,  low-brightness,  line  emission  but  no
evidence of this was found.

Tantalizingly, in the frequency  setting covering the ``blue'' profile
of the CO  J = 4--3 line some  faint emission can be seen  in the line
map,   located   between    the   two   radio   continuum   components
(Fig.~\ref{COblue}). However,  this is not  statistically significant,
especially  since the noise  measured within  the area  containing the
radio galaxy  (and where the  {\tt clean} components of  its continuum
are derived) is somewhat higher than  in the rest of the image, namely
$\sigma _{\rm eff}=\rm 65\ \mu Jy\ beam^{-1}$.  This is expected since
subtracting the  noisy {\tt clean} components,  $N_{\rm CL}$, defining
the  radio continuum  emission  will add  noise  to that  part of  the
resulting continuum-free image  ($\sigma _{\rm eff}<\rm \sqrt{2}\sigma
\sim  80\  \mu  Jy\  beam^{-1}$,  simply  because  $N_{\rm  CL}<N_{\rm
pixels}$). We  use $\sigma_{\rm eff}$ to express  the 1-$\sigma$ upper
limit of the velocity-integrated CO J = 1--0 line from

\begin{equation}
\rm \Delta  S_{1-0}=\int _{\Delta v}  \delta S_{\nu }dV = c \left( \frac{\Delta \nu
_{IF}}{\nu _{obs}} \right) \sqrt{N_{b}} \sigma _{eff},
\end{equation}

\noindent
where $\rm N_{b}$ is the  number of beams associated with the putative
CO-emitting area.  A single IF covers most of each of the detected J =
4--3   ``blue''  and   ``red''  velocity   components,   with  $\Delta
S_{1-0}=0.035$  Jy\,km\,s$  ^{-1}$  (point  source  sensitivity,  $\rm
N_b=1$),  and  a corresponding  limit  on  the  J=1--0 luminosity  per
component  of  $\rm L_{co}\sim  10^{10}\  K\,km\,s^{-1}\,pc^2 $.   The
corresponding lower limit  for the velocity/area-averaged $\rm R_{43}$
line ratios can be found from

\begin{equation}
\rm R_{43}=\frac{\langle T_b(4-3) \rangle}{\langle T_b(1-0) \rangle}=
\left(\frac{\nu _{10}}{\nu _{43}}\right)^2 \frac{S_{4-3}}{S_{1-0}}.
\end{equation}

\noindent
The  CO  J  =  4--3  emitting  regions  are  unresolved  by  the  $\rm
\theta_{HPBW}\sim  6''$ of  the Plateau  de Bure  Interferometer, thus
$\rm N_b(max)\sim  3.5$.  Using the  numbers reported by De  Breuck et
al.\ for $S_{\rm  4-3}$, and considering a 2-$\sigma$  upper limit for
the  J  = 1--0  emission,  with  $\rm  N_b=3.5$, yields  $\rm  R^{(b)}
_{43}\ga 0.30$  and $\rm R^{(r)} _{43}\ga 0.55$  for the ``blue'' and
``red'' component, the latter containing most of the H$_2$ gas mass in
this system.   For a  CO emitting region  smaller than  our resolution
limit of $\sim 3''$, $\rm  N_b=1$ and these limits become $\rm R^{(b)}
_{43}\ga 0.55$  and $\rm R^{(r)}  _{43}\ga 1$.  These are  the highest
minimum values  compatible with  the sensitivity of  our measurements,
and  the most  likely ones  since in this object $\theta _{\rm  co}\la
0.3''$ \citep{deb05}, and are adopted in all our subsequent analysis.

\section{Discussion}

The  models of the  molecular gas  physical conditions  constrained by
single  or  even  multiple  line ratios  \citep[\eg][]{aal95}  contain
significant  degeneracies, but  for modest-to-high  values of  CO line
ratios  with large  J-level differences  (and thus  difference between
$\rm  n_{cr}$ and  $\rm  E_{\rm upper}/k_B$)  these are  significantly
reduced.   The CO  J =  4--3 transition  in particular  marks  a broad
excitation turnover beyond which line flux densities and thus the line
ratios $\rm R_{J+1\, J}$ ($\rm  J+1\geq 4$) become highly dependent on
the ambient gas  conditions and can vary widely  (Nieten et al., 1999;
Carilli et al., 2002).  Moreover, because of the difficulty of routine
observations  at  $\nu \ga  \rm  \nu  _{43}\sim  461~GHz$ through  the
Earth's atmosphere,  ratios like $\rm R_{43}$ are  sensitive probes of
gas excitation that are rarely available in the local Universe.

  The lower  limit of  $\rm R_{43}\geq 0.55$  of the  ``blue'' emission
component in 4C\,41.17 is within  the range of the $\rm R_{43}$ values
measured in  another powerful radio  galaxy, 4C\,60.07 at  z$\sim $3.8
\citep{gre04},  and suggests  highly excited  gas. By  comparison, the
{\it globaly  averaged} CO  emission for the  Galaxy measured  by {\it
COBE}, yields  $\rm R_{43}\sim 0.10-0.17 $, indicative  of a quiescent
gas phase \citep{fix99}.  A single-phase Large Velocity Gradient (LVG)
model, restricted  by setting $\rm  T_{k}=T_{dust}\sim (50-60)$\,K (De
Breuck et  al., 2005), yields  $\rm n(H_2)\ga 10^3\ cm^{-3}$  for most
solutions, which  is on the upper  end of the {\it  mean} densities of
Giant Molecular  Clouds in the  Galactic disk ($  \sim\rm (10^2-10^3)\
cm^{-3}$).   If  an ensemble  of  virialized  clouds  is assumed,  the
average  cloud density  and its  associated velocity  gradient  are no
longer independent but related as

\begin{equation}
\rm \left(\frac{\rm dV}{dR}\right)_{VIR} \approx 0.65\alpha ^{1/2}\
 \left(\frac{n}{10^3\  cm^{-3}}\right)^{1/2}\ km \ s^{-1}\ pc^{-1},
\end{equation}

\noindent
where $\alpha= 0.5-2.5$ depends primarily on the cloud density profile
\citep{pap99}.   For $\rm  dV/dR =\left(dV/dR\right)_{VIR}  $  (and an
abundance  $\rm  [   ^{12}CO/H_2]\sim  10^{-4}$),  the  LVG  solutions
compatible with  $\rm R_{43}\ga  0.55$ have $\rm  T_{k}\sim (40-50)$~K
for densities  of $\rm \sim  (1-3)\times 10^3\ cm^{-3}$.   However, in
starburst  environments highly  non-virial motions  have  been deduced
\citep{dow98} and  thus LVG solution  constraints based on  known dust
temperatures   are   better   than   assumptions   of   virialized   $
^{12}$CO-emitting clouds.   For the ``red''  kinematic component, with
$\rm R_{43}\ga 1$, our models yield $\rm n(H_2)\ga 10^4\ cm^{-3}$ (for
$\rm  T_k\sim (50-60)\,  K$), typical  for H$_2$  gas  in star-forming
regions.   The  solutions  compatible  with $\rm  R_{43}>1$  and  $\rm
T_{k}\sim (50-60)\,K$ typicaly have  moderate CO J=1--0 optical depths
($\tau _{10}\sim 1-2$), a condition encountered often for the warm and
turbulent molecular gas in starburst environments~\citep{aal95}.

\subsection{The dominance of high-excitation gas in 4C\,41.17}

One-phase LVG modeling of  CO line ratios encompassing entire galaxies
(the case for  most high redshift CO line  detections), can yield only
indicative results  regarding which conditions are  prevelant in their
large  molecular gas  reservoirs.  A  different way  to  determine the
relative prevalence of  a highly-excited, starburst-fueling, gas phase
is to make use of its well-established characteristics (and typical CO
line ratios) along with those of a quiescent, star-formation-idle one.
These two phases  have been extensively studied in  the local Universe
where observations of several molecular and atomic lines, as well as a
high spatial  resolution, allow their detailed  and realistic modeling
(see Papadopoulos, Thi, \& Viti 2004 for a review).

 In the local Universe,  the unique association of $\rm R_{J+1\,J}\sim
1  $, $\rm  J+1\leq 4$  ratios (typical  of  high-excitation molecular
gas) with sites of vigorous  star formation has become apparent since
the first extragalactic detection of CO J=4--3 emission by \cite{gu96}
towards the  star-forming regions  of the nearby  archetypal starburst
galaxies M\,82,  NGC\,253 and  IC\,342 where typical  (4--3)/(2--1) CO
line ratios  of $\rm R_{43/21}\ga  0.5-0.8$ were measured.   For M\,82
this was confirmed by  much more detailed studies \citep{mao00}, while
a  recent study demonstrated  the existence  of an  additional massive
low-excitation gas  phase extending much  further ($\sim 3$  kpc) than
the starburst-related gas residing in its inner 500 pc \citep{wei05}.

For  starbursts at high  redshifts, the  distribution of  star forming
 regions is  one of the  crucial aspects that can  distinguish between
 scaled-up  (in  terms  of  star  formation rate)  versions  of  local
 starburst  galaxies (usually with  compact star-forming  regions) and
 very different,  galaxy-wide events.  Extended CO  line emission with
 large   (high-J)/(low-J)  intensity  ratios   can  be   an  excellent
 ``marker'' of  the spatial distribution of star  formation in distant
 dust-obscured  starbursts, {\it  unaffected by  extinction.}   In the
 case  of 4C\,41.17  the CO  J=4--3  emission is  unresolved, but  the
 distribution of  a high-excitation,  star-forming, gas phase  and its
 concomitant starburst  relative to the total CO-emitting  area can be
 constrained from our limit on the global $\rm R_{43}$ ratio.  Indeed,
 for  the  large velocity  gradients  characterizing molecular  clouds
 (thus  no significant radiative  ``shadowing'' between  different gas
 phases) it is

\begin{equation}
\rm R_{43} = \frac{R^{(l)} _{43} + \rho _{hl}
 f_{hl} R^{(h)} _{43}}{1+\rho _{hl} f_{hl}},
\end{equation}

\noindent
where   $\rm  R^{(h)}  _{43}$   and  $\rm   R^{(l)}  _{43}$   are  the
(4--3)/(1--0) ratios of the velocity/area-averaged, CO emission of the
high-excitation   and  low-excitation   molecular   gas  phase,   $\rm
f_{hl}=\Omega  _h/\Omega _l$ the  ratio of  their emitting  areas, and
$\rm  \rho _{hl}=[J(T^{(h)}  _k)-J(T_{bg})]/[J(T^{(c)} _k)-J(T_{bg})]$
(for  optically  thick  and  thermalized J=1--0  transition  for  both
phases), where  $\rm J(T)=h\nu/k_B (e^{h\nu/k_B  T_k}-1)^{-1}$ ($\nu =
\nu _{10}\sim  115$~GHz) and $\rm  T_{bg}=(1+z) T_{cmb} \sim 13  K$ is
the CMB temperature at $\rm z\sim 3.8$.  Setting $\rm R^{(h)} _{43}=1$
(typical for the high-excitation,  star-forming gas phase e.g.  Wei\ss
\ et al.,  2005 and references therein), and  $\rm R^{(l)} _{43}=0.15$
for the  quiescent and star-forming-idle one  \citep{fix99}, the lower
limit $\rm R_{43}\sim 0.55$  corresponds to $\rm \rho _{hl} f_{hl}\sim
0.9$.   For typical  temperatures  usually associated  with these  gas
phases: $\rm T^{(l)} _k\sim 20\,K$  and $\rm T^{(h)} _k\sim 60\,K$, it
is  $\rm  f_{hl}\sim  0.13  $.   The  latter amounts  to  $\rm  f_h  =
f_{hl}/(1+f_{hl})\sim 0.10 $ of  the total CO-emitting region occupied
by  the high-excitation gas  phase and  its concomitant  starburst (by
comparison in M\,82:  $\rm f_h \sim 0.03$).  A  maximum value for $\rm
f_{hl}$  and $\rm f_h$  can be  found by  setting $\rm  T^{(h)} _k\sim
T^{(w)} _k$ (thus $\rm \rho  _{hl}\sim 1$), i.e. the difference in the
$\rm R_{43}$  ratio (and the  excitation state of J=4--3)  between the
two gas  phases is solely due  to a difference  in densities, yielding
$\rm  f_{hl}\sim  0.9$  and  $\rm  f_h\sim 0.47$.   In  practice  both
temperature and  density effects  play a role  and the true  $\rm f_h$
will  lie  between the  aforementioned  estimated  values.  These  are
applicable to the ``blue''  component; in the ``red'' component, which
hosts  the radio-loud  AGN,  the  $\rm R_{43}\ga  1$  limit implies  a
high-excitation gas phase that fully dominates the CO emission.

\section{Conclusions}

We report sensitive new observations of  the CO J = 1--0 transition in
the distant, powerful radio galaxy  4C\,41.17 at $z=\rm 3.8$ using the
NRAO Very Large Array. Our results can be summarized as follows:

\begin{enumerate}
\item
The CO J=1--0  line emission remains undetected but  the new sensitive
upper  limit  and  the   detected  J=4--3  line  yields  a  brightness
temperature ratio of $\rm R_{43}\sim 0.55-\ga 1$ for the {\it bulk} of
its H$_2$ gas  reservoir.  This ratio is $\sim  3-6$ times higher than
the one  in quiescent  environments of the  Galaxy, and is  typical of
starburst environments.

\item   Single-phase  Large  Velocity   Gradient  models,   under  the
restriction of  $\rm T_k= T_{dust} $, yield  $\rm n(H_2)\sim (10^3-\ga
10^4)$ cm$ ^{-3}$, more typical of the high-excitation gas ``fueling''
local starbursts rather than a more diffuse, low-excitation, phase.

\item Assuming the molecular gas  reservoir to be segregated between a
high-excitation star-forming, and a quiescent star-forming-idle phase,
with $\rm  R_{43}$ ratios and  gas temperatures typical for  the local
Universe, we find the former to  occupy at least $\sim 10\%$ and up to
$\sim 100\% $ of the total CO emitting area.  This is much higher than
what  is  found  in  local  starbursts like  M\,82,  and  equivalently
demonstrates the  predominance of a high-excitation  gas phase fueling
an extreme star formation event in this distant powerful radio galaxy.

\end{enumerate}

\begin{acknowledgements}
We would like to thank James  Graham and Wil van Breugel for providing
us with  the 2.2\,$\mu$m  image of 4C\,41.17.  We are greatful  to the
anonymous referee for helping us to significantly improve  the original
manuscript.
\end{acknowledgements}

\end{document}